\documentstyle[12pt,epsfig,cite]{article}

\begin{document}
\begin{center}
{\huge Wave Dynamical Chaos in\\
Superconducting Microwave Billiards \\}
\vspace{0.5cm}
H. Rehfeld $^{(1)}$\\ and \\
H. Alt $^{(1)}$, C. Dembowski $^{(1)}$, H.-D. Gr\"af $^{(1)}$,
R. Hofferbert $^{(1)}$,\\ H. Lengeler $^{(2)}$, A. Richter $^{(1)}$\\
\vspace{0.3cm}
\small{$^{(1)}$ Institut f\"ur Kernphysik, TH Darmstadt, 
D-64289 Darmstadt, Germany}\\
\small{$^{(2)}$ Accelerator Technology Division, CERN, CH-1211 Geneva 23, 
Switzerland}\\
\vspace{0.3cm}
\footnotesize{\today}                             
\vspace {0.5cm}
\begin {abstract}
During the last few years we have studied the chaotic behavior of
special Euclidian geometries, so-called billiards, from the quantum or
in more general sense ``wave dynamical'' point of view \cite{prl92,pre94,prl95}.
Due to the equivalence between the stationary Schr\"odinger equation and
the classical Helmholtz equation in the two-dimensional case (plain billiards),
it is possible to simulate ``quantum chaos'' with the help of macroscopic, 
superconducting microwave cavities. Using this technique we investigated 
spectra of three billiards from the family of Pascal's Snails 
(Robnik-Billiards) with a different chaoticity in each case in order to test
predictions of standard stochastical models for classical chaotic systems.
\end {abstract}
\end{center}

\section{Introduction}
\label{sec_intro}

In the last few decades the theoretical investigation of two-dimensional 
Euclidian and Riemannian geometries, so-called billiards, has led to a very
fruitful new discipline in nonlinear physics \cite{birkhoff,buni,sinai}.
Due to the conserved energy of an ideal particle propagating inside
the billiard's boundaries with specular reflections on the walls, the
plain billiard belongs to the class of Hamiltonian systems with the lowest
degree of freedom in which chaos can occur. This does only depend on the 
given boundary shape. Because of their simplicity two-dimensional billiards are
in particular suitable to study the behavior of the particle in the 
corresponding quantum regime \cite{mcd,bohigas,steiner} where spectral 
properties are completely described by the stationary Schr\"odinger equation 
inside the domain with Dirichlet boundary conditions on the walls.
In this context the investigation of ``quantum chaos'' has become
one of the most fascinating goals of theoretical physics at the end of this 
century \cite{berry87,gutz}. 

The family of billiards described in this paper were introduced by M. Robnik
in 1983 \cite{robnik83} and were investigated from the classical and quantum 
point of view \cite{robnik_ges}. The shape of these so-called Robnik-Billiards
is known from mathematics as Pascal's Snails. These billiards are suited for 
an investigation of the chaotic behavior in the classical and also in the
quantum mechanical case, because by varying only one control-parameter, 
$\lambda$, the system changes from an integrable regular billiard, the circle
\cite{berry}, through a wide range of intermediate billiards to a fully 
chaotic billiard, the Cardioid \cite{baecker}. The following equation describes
the mapping from the unit disc to one certain Pascal Snail with a special 
$\lambda$: $\omega =z+\lambda z^2$ , with $|z|=1$. For $\lambda=0$ one obtains 
the circle and for $\lambda=0.5$ the Cardioid. To choose special parameters 
$\lambda$ for the investigations we calculate the Poincar\'e surface of 
section for different $\lambda$ and look for the fraction of chaotic area 
inside them. To examine the transition from the integrable to the chaotic 
case we choose the following fractions of chaotic phase space for our 
billiards: 55 \%, 66 \% and 100 \% which correspond to the following parameter
$\lambda$: 0.125, 0.15 and 0.3. Because the phase space shows no regular motion 
above $\lambda=0.279$ \cite{hayli}, we took $\lambda=0.3$ for the chaotic 
billiard.   

\section{Experiment}
\label{sec_exp}
Due to the equivalence of the stationary Schr\"odinger equation for quantum 
systems and the corresponding Helmholtz equation for electromagnetic resonators
in two dimensions, it is possible to simulate a quantum billiard of arbitrary 
shape with the help of a sufficiently flat macroscopic electromagnetic cavity 
of the same shape \cite{stock,doron,sridhar,prl92}. We have studied three 
billiards of the shape given in Sec.~\ref{sec_intro} using microwave resonators
made of Niobium which become superconducting (sc) below 9.2 K.

\begin{figure}[h]
\centerline{\epsfig{figure=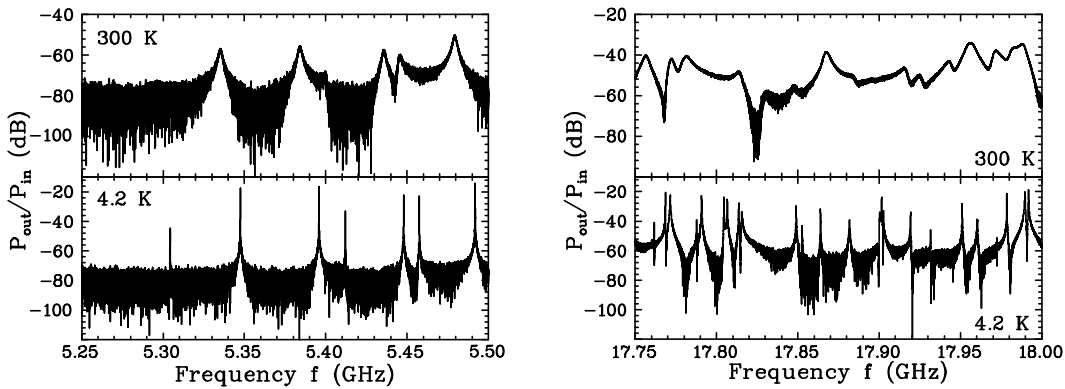,%
%width=16.5cm}}%,bbllx=3cm,bblly=3.5cm,bburx=18cm,bbury=15cm,clip=}}
width=14cm}}%,bbllx=3cm,bblly=3.5cm,bburx=18cm,bbury=15cm,clip=}}
\caption{\footnotesize{Comparision of the normal conducting (upper part, at 
300 K) and the superconducting (lower part, at 4.2 K) transmission spectrum of 
the $\lambda=0.15$ billiard for two different frequency ranges.}}
\label{spectrum}
\end{figure}

The measurements were carried out in a very stable 4K-bath-cryostat
\cite{altetal}. The billiards were excited in a frequency range between 0 and 
20 GHz. We used capacitively coupling dipole antennas placed in small holes on 
the Niobium surface. Using one antenna for the excitation and either another 
as well as the same one for the detection of the microwave signal, we were able
to measure the transmission or the reflection spectrum of the resonator by 
using an HP-8510B vector network analyzer. In the lower part of 
Fig.~\ref{spectrum}, a typical transmission spectrum at 4.2 K is shown for two 
different frequency ranges. The signal is given as the ratio of output power 
to input power on a logarithmic scale. By comparing the upper with the lower 
part of Fig.~\ref{spectrum} the advantages of using sc resonators instead of
normal conducting (nc) ones are clearly visible. Obviously the use of sc 
resonators leads to an immense improvement of detecting almost all resonances 
in the spectra. In the nc spectrum, due to the very broad resonances which 
interfere with each other, one is not able to detect all resonances in the 
upper frequency range. Only in the sc case one can be sure to find nearly 
all resonances. In the sc case resonances typically posses quality factors of 
up to $Q=f/\Delta f\approx 10^7$ \mbox{(nc $Q\approx 10^3$)} and signal-to-noise 
ratios of up to $S/N\approx 60$ dB \mbox{(nc $S/N\approx 30$ dB).} This high 
resolution allows easy  separation of individual resonances from each other and
from the background. As a consequence, all the important characteristics like 
the eigenfrequencies and widths could be extracted with avery high accuracy 
\cite{prl95,nucphys/physlettb}. A detailed analysis of the original spectra 
yielded a total number of 
about 1100 resonances for each of the three measured billiards.

\section{Results and Discussion}
\label{sec_resu}
In order to derive meaningful statistical measures for the given eigenvalue 
sequences it is first of all necessary to extract the smooth part of the 
resonator's number of eigenmodes which is given by the generalized 
electromagnetic Weyl-formula \cite{weyl,baltes}
\begin{equation}
N^{Weyl}(f)=\frac{A\pi}{c^2_0}f^2-\frac{U}{2 c_0}f+const.\ \  ,
\label{weyl}
\end{equation}
where $A$ denotes the area, $U$ the circumference of the cavity and $f$ the
upper frequency limit of the given spectrum. The constant term contains 
contributions from the boundary's curvature and from the edges of the cavity.
The total number of eigenmodes $N(f)$, up to a certain frequency $f$, i.e. the
spectral staircase function, contains in addition a fluctuating part
\begin{equation}
N(f)=N^{Weyl}(f)+N^{fluc}(f)=\sum_i\Theta (f-f_i)=\sum_{i \atop f>f_i} 1 
+\sum_{i \atop f=f_i}\frac{1}{2} \ \ .
\label{number}
\end{equation}
To determine the spectral fluctuations, the smooth part of the spectral 
staircase has to be eliminated. For this a special staircase function 
(see Eq.~(\ref{number})) was constructed and a second order polynomial, 
Eq.~(\ref{weyl}), was fitted to it.

\begin{figure}[h]
\centerline{\epsfig{figure=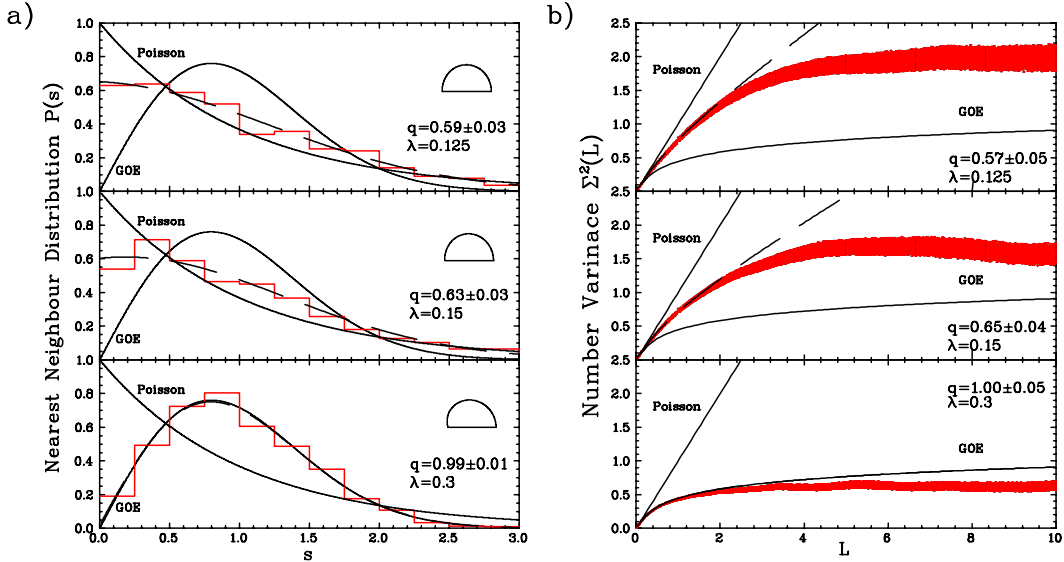,%
%width=16.5cm}}%,bbllx=3cm,bblly=3.5cm,bburx=18cm,bbury=15cm,clip=}}
width=14cm}}%,bbllx=3cm,bblly=3.5cm,bburx=18cm,bbury=15cm,clip=}}
\caption{\footnotesize{a) Nearest neighbour spacing distributions for the 
three measured configurations. The dashed lines show the best fits using the 
model of Berry-Robnik on the data, given as histograms. Also the two limiting 
cases (Poisson and GOE) are displayed. The inserts of this figure show the 
shape of the investigated billiards. b) Number variance $\Sigma^2$ for the 
three billiards. The shaded bands display the data including error bars and the
dashed curves represent the fitted Berry-Robnik distributions corresponding to
the given mixing-parameters. For the billiard with $\lambda=0.3$ the data and 
the fitted curve fall together.}}
\label{nnd_sigma}
\end{figure}

In order to perform a statistical analysis of the given eigenvalue spectra
independently of the special sizes of the resonators, the spectra were first 
unfolded i.e. from the measured sequence of eigenfrequencies
$\{f_1,...,f_i,f_{i+1},...\}$ the spacings $s_i=(f_{i+1}-f_i)/\overline{s}$
between adjacent eigenmodes were obtained by calculating the local average
$\overline{s}$ from the fit of Eq.~(\ref{weyl}). The proper normalization of 
the measured spacings of eigenmodes then yielded the desired nearest neighbour 
spacing distribution $P(s)$ (NND), the probability for a certain spacing $s$. On the 
left side of Fig.~\ref{nnd_sigma} the three nearest neighbour spacing
distributions of the measured resonators are shown.

Furthermore, to obtain a quantitative criterion concerning the degree of
chaoticity in the system the spectra were analyzed in terms of a statistical
description introduced in a model of Berry and Robnik \cite{berrob} which 
interpolates between the two limiting cases of pure Poissonian and pure GOE 
behavior for a classically regular or chaotic system respectively. The model 
introduces a mixing-parameter $q$ which is directly related to the relative 
chaotic fraction of the invariant Liouville measure of the underlying classical
phase space in which the motion takes place, see also Sec.~\ref{sec_intro}.
Using this ansatz one obtains the mixing-parameters for the three measured 
cavities: $q=0.59\pm 0.03$ for $\lambda=0.125$, $q=0.63\pm 0.03$ for 
$\lambda=0.15$ and $q=0.99\pm 0.03$ for $\lambda=0.3$.
 
To uncover correlations between nonadjacent resonances, one has to use a 
statistical measure which is sensitive on larger scales. As an example we used 
the number variance $\Sigma^2$ \cite{bohigas}. The $\Sigma^2$-statistics 
describes the variance of the number of levels $n(L)$ in a given range of 
length $L$ around the mean for this interval, which is due to the unfolding 
equal to $L$,
\begin{equation}
\Sigma^2(L)=\bigg\langle \left(n(L)-\langle 
n(L)\rangle_L\right)^2\bigg\rangle_L=
\langle n^2(L)\rangle_L-L^2 \ \ .
\end{equation}
On the right side of Fig.~\ref{nnd_sigma}, the number variances for the three
measured billiards are displayed, again together with the two limiting cases
for Poissonian and the GOE behavior. Using the Berry-Robnik model yields the 
following mixing-parameters: $q=0.57\pm 0.05$ for $\lambda=0.125$, $q=0.65\pm 
0.04$ for $\lambda=0.15$ and $q=1.00\pm 0.05$ for $\lambda=0.3$.
Comparing short range (NND) and long range correlations ($\Sigma^2$) one
observes a good agreement of the mixing parameters. The same is true if one 
compares the classical mixing-parameter from  Sec.~\ref{sec_intro} with the 
spectral ones.

\begin{figure}[h]
\centerline{\epsfig{figure=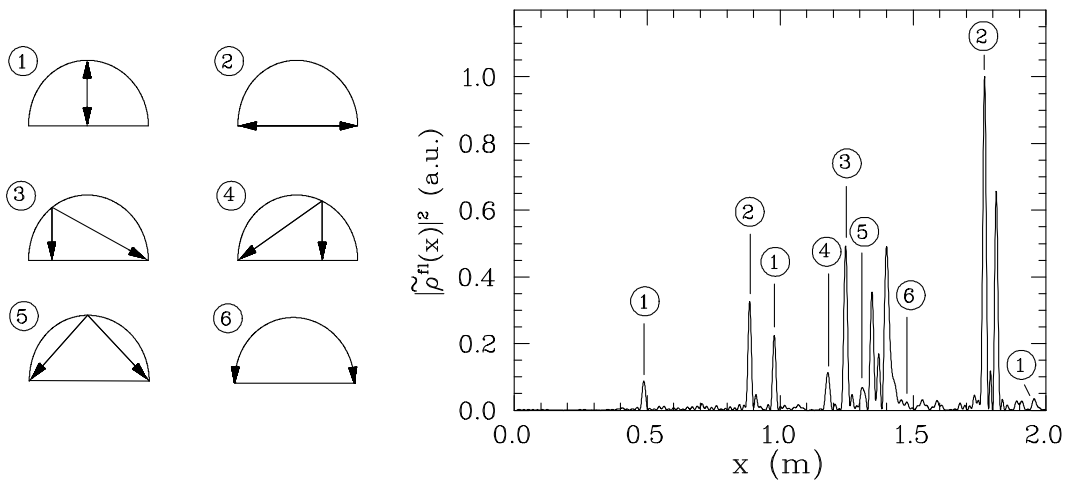,%
%width=15.5cm}}%,bbllx=3cm,bblly=3.5cm,bburx=18cm,bbury=15cm,clip=}}
width=14cm}}%,bbllx=3cm,bblly=3.5cm,bburx=18cm,bbury=15cm,clip=}}
\caption{\footnotesize{On the left side the shortest periodic orbits for the 
classical billiard ($\lambda=0.3$) are sketched. On the right side the 
Fourier Transform of the fluctuating part of the level density is shown. 
The encircled numbers above the  peaks label the classical periodic orbits.}}
\label{four}
\end{figure}

Beside the comparison between the classical and the spectral mixing-parameter,
there is another method which allows to identify relations between the 
classical and the quantum system. Taking the Fourier Transform of the 
fluctuating part of the level density, given through Eq.~(\ref{number}), one 
obtains the length spectrum of the resonator \cite{gutz}. Here the peaks 
correspond to the periodic orbits (PO) of the classical billiard. The heights 
of these peaks are a measure for their stability. On the right side of 
Fig.~\ref{four} the Fourier spectrum of the fluctuating part obtained from the 
$\lambda=0.3$ billiard is displayed. On the left side of this figure the 
shortest POs of the classical billiard are sketched and related to the length 
scale in the Fourier spectrum by labels.

We have examined the behavior of systems which are located inbetween 
integrable and chaotic characteristics. We have also shown the necessity of 
using superconducting microwave billiards for a proper examination of the wave 
dynamical system in order to get the heighest possible resolution.

We thank the collaborators at the CERN workshop for their precise manufacturing
of the niobium cavities. This work has been supported by the 
Sonderforschungsbereich 185 ``Nichtlineare Dynamik'' of the Deutsche 
Forschungsgemeinschaft and in part by the Bundesministerium f\"ur Bildung und 
Forschung under contract No. 06DA665I.

\end{document}